# NET-CENTRIC WORLD:
# LIFESTYLE OF THE 21ST CENTURY

Daniel Kharitonov

*Juniper Networks Inc., Sunnyvale CA USA*
dkh @ juniper dot net

**ABSTRACT**

*In this paper, we research the potential of information communication technologies (ICTs) for changing our society from a commute-centric to a network-centric environment. We propose to formalize the key attributes of ICT-based telecommuting experiences from both economic and human interactivity perspective. We introduce the notion of network-eligible transactions and disclose the link between degree of network centricity and worker settlement radius, postulating that media-rich network services have a strong potential to increase the physical distance between work and home locations. We also highlight notable technology challenges and opportunities of migration from location-based to mobile living, signifying the needs for new services and standards development.*

***Keywords***— ICT, net-centric, telecommuting, nomadic

## 1. INTRODUCTION

Since prehistoric times, the concentration of human activity has been synonymous with density of population. Having started as early as 3.000-4,000 BC, the continuous process of urbanization still goes on today, with the UN estimating that about half of the world is now living in metropolitan areas [1]. With large cities becoming focal points for opportunities, services, and wealth, they tend to attract the massive daily migration of humans also known as *the commute*. Commuting allows workers to reside beyond walking distance from their jobs in exchange for certain inconveniences such as unproductive time loss (averaging over 100 hours per year in the U.S. [2]), pollution of the environment, and transport expenses. While most large cities incessantly invest in mass transit infrastructures, the ongoing shift of economies in developed countries from goods to services [3] suggests the possibility that a growing percentage of commuters could, in fact, use ICT facilities in lieu of their physical presence at manufacturing worksites. And this percentage could be quite significant. A 2010 survey of U.S. government employees [4] revealed that 55% were eligible for teleworking, but only 8.67% of respondents used this opportunity. An even larger gap was found in the ICT sector, where the 2008 survey of 1,500 U.S. professionals [5] found that 37% were genuinely interested in telecommuting to such a degree that they would accept a pay cut, but only 7% could effectively work remotely. Such a discrepancy signifies that the economic and social impact of ICT has not reached its full potential in the workspace, and many aspects of networked humanity remain unidentified. In this publication, we intend to explore the mechanism of telecommuting relative to the modern state of communications technology and uncover the potential social consequences of this relation.

## 2. COMMUTE-CENTRIC WORLD

The traditional view on human behavior at work with respect to commute patterns suggests that employees have two choices—show up at the work desk or stay at home and work remotely. This is why many studies and surveys focus strictly on ecological, transport, or economical outcomes of home-work interchange.

However, it is easy to observe that relatively few individuals have an option to reside at arbitrarily selected locations and most spatial-based choices in our lives are dependent upon each other. For example, commute maps tend to strongly correlate with real estate prices (Figure 1), suggesting that the cost of housing plays a significant role in choosing places to live and work.

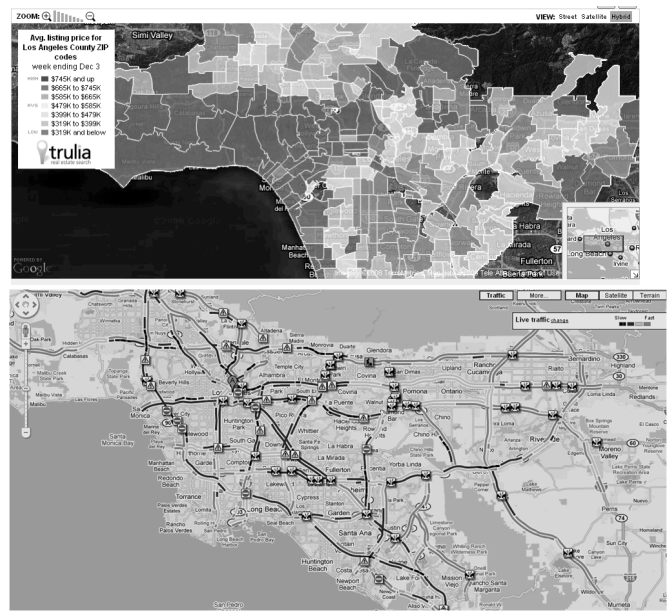

Figure 1. Residence pricing versus rush-hour commute map (source: Google, Trulia)

A graph very similar to Fig.1 is trivial to compile at any location where real-time traffic information is available: the most congested routes tend to be ones connecting the areas with notably different cost of residence.

Whenever a main commute anchor is dropped (e.g., by accepting a job offer or finding a neighborhood with good schools), matching decisions are made to maintain daily activities within a reasonable distance. This means that our lifestyle today is not significantly different from the commute-centric environment of our ancestors, who were bound by pedestrian or equestrian range. A median U.S. worker nowadays lives within 16 miles of work [6], median UK commuters settle within 5 to 10 miles [7], and Canadian workers reside on average within only 7.6 kilometers from their offices [8].

But why exactly do we need to stay physically close to colleagues, businesses, and services while living in what seems to be an increasingly information-driven society?

The answer could be in the fact that we seem to assign tangible value to nonverbal interactions. In his foundational study on how humans store and retrieve information, Edward T. Hall postulated, "language, the system most frequently used to describe culture, is by nature poorly adapted to this difficult task" [9].

The multisensory aspect of human existence was subject to numerous psychology and performance studies. In one example, research of scientist collocation found that the mean citation index for (first-last) author relationship in publications decreased as the physical distance between them increased in the three categorized ranges (same building, same city, or different city) [10]. In another example, applied psychology work discovered that high intensity telecommuting exacerbates the negative impacts on quality of interpersonal relations [11], suggesting certain workspace conflicts (at present state of technology) cannot be effectively resolved without direct human contact.

Such and similar research may shed some light on why large-scale telecommuting (as function of technology) is still not reality, even among the well qualified and eligible social groups.

This brings the logical question—are we bound to live in a commute-centric world and if not, what can be done to change that?

### 3. TAXONOMY OF HUMAN INTERACTIONS

Significant amount of academic work on cognitive engineering and individual performance suggests that humans are highly capable of multidimensional judgements, with processing in correlated dimensions (such as being able to hear and see a person) improving security of channels and reducing information loss [12].

In fact, there appears to be a broad range of highly interactive, multidimensional experiences, which underpins important functions, cues, and customs of a human society.

For example, numerous studies postulated that spontaneous and informal transactions like hallway conversations, dinners, and face-to-face brainstorms could be important to creating and maintaining productivity in the workplace [13][14]. Similarly, certain high value services (such as medical consultations and wealth management) are perceived to be more productive when exercised at considerable length in person or across a multitude of channels versus "purely electronic", mono-channel engagements [15][16].

If we adopt the view that the main value of interpersonal communications lies in richness of informational channels, the necessary conclusion should be that the online workflow (including telecommuting experiences) can be vastly improved with a transition to multichannel information exchanges (such as telepresence and virtual reality systems). This position is supported by evidence that human information-processing capabilities improve with redundant message coding across multiple modalities [17].

However, the replacement of eligible physical modalities with their virtual equivalents is not straightforward.

First of all, digital interactions are grafted over a complex web of technical appliances such as local area networks, wide area exchanges, user terminals, applications, and so forth. As a result, the quality and precision of electronic experiences can go down when details are lost "in translation" due to noise, latency, analog-to-digital conversion, compression, and other technology artifacts.

Second, certain sensory dimensions – such as olfactory, gustatory and kinesthetic experiences are hard to reproduce remotely, at least with existing telepresence equipment.

Finally, whenever we introduce the network into human-to-human transactions, the cost of delivery changes significantly.

Assuming it takes the same time to conduct an interactive session face-to-face or electronically, the main direct cost associated with "mortar-and-brick" participation is related to transport, i.e., the ability to meet a colleague, business partner, or physician in person may cost from pennies to thousands of dollars (a long-haul flight), with the U.S. average being 50 cents per mile (Table 1).

Indirect costs can run much higher. A worker may need to pay a premium for a house that is within an acceptable distance to work, an employer may have to sustain the soaring cost of office space in the middle of a good business district, and so on.

TABLE I. HUMAN-HUMAN INFORMATION EXCHANGES

| Exchange | Media | Delivery cost |
| --- | --- | --- |
| Message | Written memo or note | $0.44/letter |
| Verbal | Conversation | $0.5/mile |
| Visual | Face-to-face meeting | $0.5/mile |
| Multimodal | Lunch, hallway talk, physical treatment, brainstorm session | $0.5/mile |

On the opposite, network-based interactive transactions are priced according to "buckets" of connectivity (Table II). For instance, we may largely assume that multimodal telepresence and virtual reality (VR) systems are free of telecom charges within the corporate local area network (LAN), but they may not work over consumer-grade Internet connections such as residential broadband lines. On the other hand, running a private leased circuit from home to office for full-scale telepresence can be feasible in large metropolitan areas but priced out of reach for all but the wealthiest telecommuters.

TABLE II. HUMAN-NETWORK-HUMAN INFORMATION EXCHANGE

|  | Virtual Experiences | | | | |
| --- | --- | --- | --- | --- | --- |
|  | *Message* | *Verbal* | *Visual* | | *Rich multimodal* |
| **Media** | IM/SMS | Phone | Video Stream | Telepresence | Virtual Reality, 3D Video, etc. |
| Quality/reliability | Medium | High | Low | High | High |
| Minimum bandwidth | 160 Bytes | 9 Kbps | 0.2-2 Mbps | 2-4 Mbps | 5-20 Mbps |
| QoS requirements | Low | High | Medium | High | Very high |
| **Proximity/Cost** | | | | | |
| Local/LAN | $0 | $0 | $0 | $0 | $0 |
| Metro area/DSL | ~$0 | ~$0 (VoIP) | ~$0 (IP Video) | $0.5/min[c] | $10/min[c] |
| National mobile | $0.1[a] | $0.25/min[a] | $0.05/min[b] | Not supported | Not supported |
| International mobile | $0.5[a] | $4[a] | $10/min[b] | Not supported | Not supported |

a) U.S. average GSM voice and short text tariffs  b) U.S. 3G data tariffs  c) U.S. leased line tariffs based on one hour/day usage

Further, as employees leave their residential areas and work on the go, their mobile carrier will also charge them for airtime. A cellular service provider will typically bill for all messages, calls, and data transactions conducted by smartphones and wireless capable tablets, while frequently imposing restrictions on media experiences (audio, video) that may become crippled in quality or accessibility.

Finally, telecom expenses can mount fast outside the user's home country, as crossing the border often invites roaming fees. It is still not uncommon for smartphone users to face abnormally high telecom charges accumulated abroad over rather trivial usage profiles [18].

## 4. NET-CENTRICITY

Assuming employers are generally willing to support remote collaboration, we can now formulate a hypothesis that economy of telecommuting is primarily driven by the confluence of available interaction levels and related costs. An act as simple as reconciliation of a business discussion may be impossible without a trip to meet one's peers and shake hands; at the same time, a transaction as complex as surgery can be successfully done remotely despite extremely high telecom and robotic equipment costs [19].

Thus, we can presuppose that telecommuting is only practical when it offers a suitable compromise between the cost of information exchanges and the ability of ICT infrastructure to sustain an effective workflow. If achieved, such compromise should mark an important change in human behavior—the increasingly connected work ecosystem becomes less dependent on physical distances (commute) but more dependent on availability, economy, and quality of ICT services (network). Since the price and availability of network services are largely decoupled from their physical location, we can reasonably claim that this new behavior model starts to drift away from legacy, location-driven society. To reflect this difference, we will refer to the new, nomadic lifestyle as "net-centric."

### 4.1. Net-Centric Factor

The first question that comes to mind when defining characteristics of a net-centric world is what percentage of duties can be fully performed over existing telecom infrastructure. In the pre-Internet era, very few occupations were eligible for full-time telecommute, with the rest of the economically active population glued to physical workplaces. Today, a significant percentage of the population in developed countries may, in fact, perform duties remotely, at least partially [20]. Thus, every job can be described with a metric that reflects the percentage of work that can be robustly and economically done over the network. Let's call such a metric a net-centric factor (NCF):

**NCF** = online tasks / (offline tasks + online tasks)     (1)

For example, a professional technical writer, who does not depend on personal collaboration with co-authors or publishers, may achieve an NCF close to one. On the other hand, a hair stylist will likely have an NCF = 0 simply because specialized machinery for remote coiffeur services is economically prohibitive to build, given the prevailing haircut rates. Considering that every active worker may have a unique combination of possible online and offline actions and duties, NCF is highly personalized. For example, a person who may effectively come to the office three days a week has a de facto NCF factor of 0.4 (forty percent tasks can be done offline), while a peer in the same job may have an NCF factor of 0.1 or even less[1].

---

[1] Although there is some evidence that employee output may change by the mere act of telecommuting or fluctuate with tenure, skill or task interdependency [21][22], in this paper we assume that at any career point, their NCF can empirically established.

It is also important to note that NCF merely describes the potential for effectively doing the job remotely and has to be augmented by availability and cost of technology. From a practical standpoint, NCF denotes the minimal frequency of commute required to maintain normal productivity level at work. An NCF of 0.6, for example, allows for commute twice a week, an NCF of 0.8 once a week, and so on. Also of note is that high NCF values do not necessarily mean a proportional reduction in transport distances or expenses, as a teleworker may come to the office more frequently or choose to reside further away from it.

**4.2. Net-Centric Economy**

When discussing the taxonomy of human interactions, we have mentioned that the cost of in-person transactions is linear and consists of commute expenses plus an indirect premium for residing within an acceptable commute radius. On the other hand, the cost of pure networked transactions is discrete and is entirely driven by connectivity. Thus, the combined economy of living in the net-centric world can be formalized with this equation:

$$Sv > \sum_{i=1..N} C_i t + C_c * (1 - NCF) \quad (2)$$

where

$Sv$ denotes the value of remote work due to better location, cheaper housing, better living conditions, etc.

$Sum_i (C_i t)$ denotes the sum of telecom transactions across all N media types required to support online tasks. This includes amortization cost of all necessary software programs and hardware appliances.

$C_c$ denotes the cost of daily commute, including transportation, security, time, and other expenses needed to support offline tasks.

One of the possible ways to quantify $Sv$ is to note that the cost of real estate is typically inversely proportional to a settlement radius. For instance, if an employee works in San Francisco's financial district, the cumulative cost of housing and transport (based on a five day work week) may look similar to that shown in Table III.

When we calculate housing costs based on a typical estate pattern similar to that shown in Figure 1, the immediate vicinity of premium office space (0–5 miles) commands the highest prices, which gradually decrease as the residence moves away from the business center and into suburbs. At the same time, commute costs build up both with distance and the number of commute days per week. This explains the empirical "sweet spot" found by surveys—without telecommuting (NCF = 0), it is most economical to settle within a (certain) city, country, and region-dependent optimal distance from work (U.S. average being 16 miles).

However, as NCF increases, so does effective settlement radius. Working remotely two days a week (NCF = 0.4) makes it feasible to reside a bit closer to work, but also strongly motivates workers to move further away from the office. Shown in bold in Table III, the acceptable telecommuting solutions (cost of housing plus transport less or equal to that of the best location with NCF=0) clearly demonstrate that the settlement radius increase is proportional to NCF.

Higher NCF values may also result in new lifestyle options. A high-intensity telecommuter coming to the office once a week (NCF = 0.8) can economically reside within regional jet or a high-speed train reach and be qualified for such work-home combinations as "San Francisco–San Diego" or "Zurich-Berlin." Additionally, the equation (2) suggests even cross-continental work habits may make economical sense. With an NCF of 0.8 or more, an employee may live away from Northern California to as far as Hawaii (2,400 miles) or Montreal (2,600 miles) - considering the difference in median house prices, this may actually make a case for relocation. Even more surprisingly, the carbon footprint of cross-continental commuters can still beat their office-dwelling colleagues. With the U.S. national average of 15,000 miles per driver, any employee using a car with ordinary fuel efficiency could just as well spend about 50 hours per year on commuter jets.

TABLE III. SAMPLE ECONOMIC IMPACT OF NCF

| Distance to Work/ Time to Work | House/ Month | Commute Mode | Cost* | Monthly Cost of Housing Plus Transport | | | | | | |
|---|---|---|---|---|---|---|---|---|---|---|
| | | | | NCF = 0 | 0.2 | 0.4 | 0.6 | 0.8 | 0.9 | 0.95 |
| 0 miles / 5 minutes | $5,000 | Walk | *$0* | *$5,000* | *$5,000* | *$5,000* | *$5,000* | *$5,000* | *$5,000* | *$5,000* |
| 5 miles / 20 minutes | $3,500 | Tram/Bus | *$25* | $4,000 | $3,900 | $3,800 | $3,700 | $3,600 | $3,550 | $3,525 |
| 10 miles / 30 minutes | $2,500 | Car | *$40* | $3,300 | $3,140 | **$2,980** | **$2,820** | **$2,660** | **$2,580** | **$2,540** |
| 25 miles / 45 minutes | $1,500 | Car | *$71* | **$2,980** | **$2,636** | **$2,352** | **$2,068** | **$1,784** | **$1,642** | **$1,571** |
| 40 miles / 60 minutes | $1,200 | Car | *$100* | $3,200 | $2,800 | **$2,400** | **$2,000** | **$1,600** | **$1,400** | **$1,300** |
| 100 miles / 120 min. | $1,100 | Car | *$200* | $5,500 | $4,620 | $3,740 | **$2,860** | **$1,980** | **$1,540** | **$1,320** |
| 1000 miles / 180 min. | $900 | Train/Air | *$480* | N/A | $6,980 | $5,460 | $3,940 | **$2,420** | **$1,660** | **$1,280** |
| 2500 miles / 300 min. | $1,100 | Air | *$900* | N/A | N/A | $12K | $8,300 | $4,700 | **$2,900** | **$2,000** |
| 2500 miles / 300 min. | $1,200 | Air + hotel[†] | *$900* | N/A | N/A | $5,400 | $4,100 | **$2,900** | **$2,500** | **$1,850** |
| 6000 miles / 840 min. | $1,300 | Air + hotel[†] | *$2,340* | N/A | N/A | N/A | $7,580 | $4,440 | **$2,870** | **$2,085** |

\* Roundtrip cost, including time loss at $0.5/minute and transport at $0.5/car mile or $200/$600/$1,500 for short/mid/long-haul airtickets
[†] Housing cost includes $200/night hotel surcharge on commute days.

Finally, nomadic and ultra long-haul commutes make an extreme, but still a sound business case. With the ability to reside at the worksite temporarily (e.g., using hotels for accommodation), telecommuting may span hemispheres.

In this latter case, the cost of housing should not be the only (or most important) reason for living at remote locations, so Table III assumes that ultra long-haul commuters pay more than the lowest neighborhood prices for their choice of residence and ability to stay close to transport hubs.

If we plot Table III as a function of cost of living relative to commute distance, we will observe that all existing workers residing at a nontrivial distance from the office can financially benefit from an increased amount of telecommuting (Figure 2).

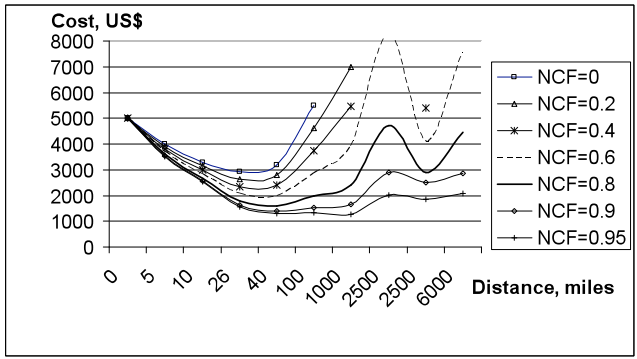

Figure 2. Cost of housing and transport relative to commute distance for various NCF values

This forms the basis for the ICT business case. If telecom services can act as enablers for higher intensity telecommute, consumers and service providers can benefit together from a transition to the new work model.

## 5. ICT COSTS AND OPPORTUNITIES

So far we have intentionally stayed away from quantifying the expression $Sum_i (C_i t)$, i.e., we did not put any bounds or restrictions on cost of telecom services required to support high-intensity telecommuting.

Such boundaries can be trivially established by resolving expression (2) for known values of Sv, NCF, and transport. If we plot the cumulative financial gain from telecommuting due to lowering transport expenses using sample data from Table III (Figure 3), we notice that workers residing within immediate vicinity to work (0-5 miles) do not have financial drivers to practice low intensity telecommuting (NCF 0.2 to 0.4). This category of workers may still realize some savings (less than $500 per month) from higher intensity telecommuting patterns, but they are not likely to be motivated to increase spending on telecom products beyond their normal utility packages.

Quite predictably, workers with residences beyond the average distance may realize sizeable profit from even low-intensity telework patterns - such as coming to the office three to four times a week. These people should be financially interested in sustaining their net-centric lifestyle, as they can definitely increase their telecom spending beyond the minimum package and occasionally may afford services with recurring monthly costs up to $500 (or even higher). A typical user from this group would be an executive or highly paid professional whose telecom expenses can be partially covered by the company or may remain insignificant relative to salary.

However, the most interesting case is seen at the median commuter radius (10-40 miles), where cost-conscious workers may gain $200 to $500 per month with minimal telecommuting efforts and up to $1,000 or more for higher intensity telecommuting.

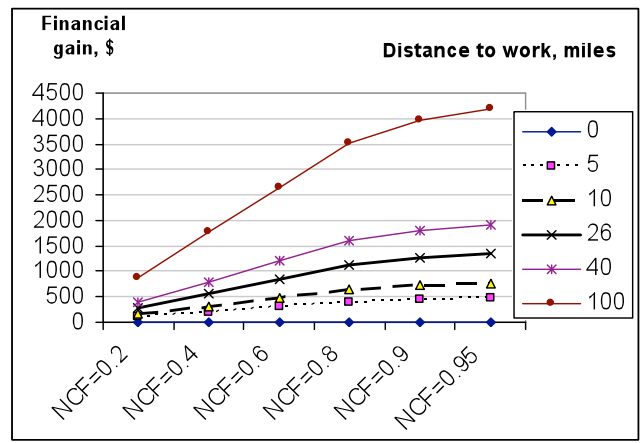

Figure 3. Financial gain due to reduction in cost of housing and transport relative to commute distance

This latter group of customers makes the "bread-and-butter" business case—if telecom providers can provide quality work-home communications at residential locations, they should be able to increase monthly account charges by $50 to $200 – in other words, nearly double or triple the current average revenue per user (ARPU).

However, matching the content of Table II against this target revenue also reveals a paradox—the current telecom industry does not provide products that can enable multimodal experiences in a suitable price range.

This gap is surprising, because the ability of humans to absorb information across sensory channels is biologically constrained [23][24] while the progress of codecs, presentation, and broadband access speeds continues at a steady rate. In fact, we can postulate that the ability of ICT systems to transport the amount of information matching the capacity of all human sensory channels is unquestionable, with the only problem being how to cross this barrier economically.

Moreover, the majority of urban population in developed countries already has access to broadband Internet in the speed ranges quite suitable for high quality video streaming [25], while a growing percentage of fiber-connected residents may afford to run virtually any available streaming application. According to Organization for Economic Cooperation and Development (OECD), the median advertised broadband download speed in 2010 was 15 Mbps with prices ranging from $0.13 (Japan) to $11 (Mexico) per megabit per second [26]. Therefore, in theory, rich multimedia experiences can be economically delivered in most significant urban hubs worldwide.

In reality, however, Internet service providers (ISPs) remain mostly oriented towards best-effort services, with no guaranteed connections (virtual leased lines) offered to consumers even within their "home" network much less across different service providers.

For instance, a popular voice over IP (VoIP) and video conferencing applications Skype uses an array of audio codecs, including G.729 with lowest bitrate of 8Kbps [27]. At the same time, business-grade VoIP platform Skype Connect™ manual recommends the minimum of symmetrical 33Kbps connection speed (up to six sessions over 256Kbps/512Kbps ADSL service), suggesting 4x the bandwidth over-provisioning to cover for lack of explicit QoS on the Internet connections users [28].

In another example, a leading US streaming provider Netflix reveals that their subscribers on top US networks are able to watch TV and movies at speeds ranging from 1400 Kbits per second to 2700 Kbits per second, with "no client being able to sustain 4800 <Kbits per second> stream from start to finish" [29]. Considering the fastest service provider from Netflix list (Cablevision) in 2011 offered the minimum download access speed of 15Mbps, it took over 5x of over-provisioning to maintain one video streaming application. It is even more interesting to note, that Netflix application typically runs between directly connected networks - last-mile Internet service provider and content-delivery operator like L3, Limelight or Akamai.

So practically speaking, while ordering targeted QoS parameters from any broadband provider today is not possible, consumers and businesses have to pay for access speeds several times higher than the bitrate required for applications they are interested in.

By extension of this example, if we consider running rich media session with over Internet in business environment with quality parameters similar to that of needed by Skype (0.2% or less packet loss, 10ms or less jitter and 200ms or less of delay), a broadband connection required to support telepresence or virtual office sessions may need sustainable access speed ranging from 40 to 200 Mbps—something not feasible in the nearest future, especially over copper or airwaves.

## 6. CALL FOR STANDARDIZATION

In the previous section we hinted at the possibility of new, high margin telecommunication services to support interactive applications. For example, a high-speed, guaranteed QoS "virtual leased line" between home and office might, in fact, become a popular service if priced to satisfy the restrictions of our equation (2). We can also foresee a market segment for novel types of consumer collaboration and media applications such as virtual offices, virtual multimodal meeting rooms, and so on.

However, the task of developing signaling, forwarding, and billing solutions for inter-provider QoS-aware tunnels in a generalized, N-service provider format (Figure 4) presents a notable challenge for vendors, network architects, and standards organizations alike.

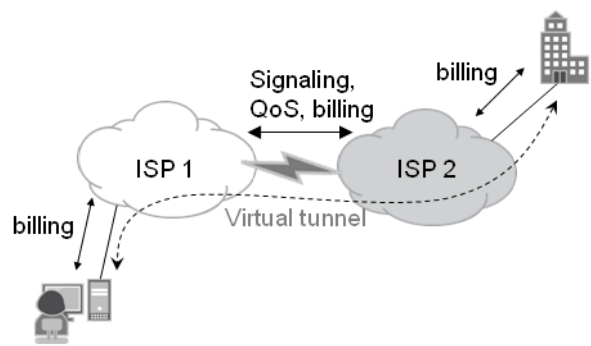

Figure 4. Virtual tunnel between home and office

On one hand, service providers are hesitant or unwilling to invest into developing proprietary application architectures that may not find a matching interface at their Internet peers. Therefore, early involvement of international standards organizations (such as ITU-T) is a must.

On another hand, standardization of all-inclusive network orchestration architectures is a slow, daunting task that requires designing consistent and unified policy management in a system with complex moving parts. The partial list of problems to solve includes mechanisms for connection admission and control, resource reservation, runtime verification of service-level agreements (SLAs), multiparty billing and packet handling programming for transit network devices (routers and switches). Last (but not least) are security concerns that include control for misconfigurations, runaway client devices and network resource abuse by humans and robots as well as integrity and confidentiality guarantees for client data.

This is why, despite the strong body of work on external network-to-network interface (E-NNI) definitions by various organizations including Optical Internetworking Forum [30], Metro Ethernet Forum [31] and pioneering efforts by IPSphere framework group within Traffic Management Forum [32], practical development of session-based inter-provider QoS services and interfaces remains in the early stage.

The relatively poor condition of de facto and de jure standardization in this area can also be (partially) explained by scarcity of session-based IP services suitable for immediate monetization. However, this deficit works both ways—the lack of session-based services is also an inhibiting factor for development of novel interfaces, interactions, and learning technologies. Thus, fostering and encouraging standardization efforts in this area should resume from making a clear, executable mission statement on the technical subject and related business case.

## 6. STANDARDIZATION PROPOSAL

The concept of net-centricity assumes close, robust relation between remote workers and media-rich corporate services over network infrastructure. We consider this concept to be pivotal to a future stack of standards defining provider-agnostic application-aware networking (AAN) [33].

If we take a closer look at Figure 4, we may notice it can be simplified into two possible architectures – (1) Content platform based interactive services and (2) "walled-garden" based interactive services.

In a first use-case, a corporation willing to offer rich media experience to remote workers moves its content (such as virtual office environment or 3D telepresence sessions) to a commercial content delivery network (CDN). Considering that CDNs maintain direct peering with all major service providers, this move guarantees that QoS planning, delivery and external network-to-network interfacing remains constrained within the latter, thus greatly simplifying the original IPSphere service planning model [32]. Once last-mile service provider authorizes and accepts service request from the user, it is routed to the nearest CDN operator, which in turn bills content owner based on usage. In that case, access operator acts as both Element Owner (participating in cross-domain design and delivery) and Administrative Owner (offering its own transport services for retail). This allows for session admission and control to run only once (at customer interface) and billing to be complete in two cycles (CDN to corporation and ISP to CDN), while reliably serving the needs of large national and international businesses and their remote employees.

In a second use case, a corporation willing to offer rich media experiences to remote workers moves its content platform directly into "walled service garden" of the last-mile service provider. This model reduces the number of parties to two, but has disadvantage of lower scaling parameters (one content platform is needed per every supported ISP) and better fits regional businesses.

Both architectures are significantly simplified relative to all-encompassing QoS architecture that is required to support an arbitrary number of applications over chain of service providers with complex mix of capabilities. On the other hand, our proposal can be seen as stepping stone for evolved application-aware services – such as subscription-based gaming, remote medical diagnostics, 3D webcasts and others.

The implications of proposed standardization efforts can be significant.

With 90% of the world's metro areas already residing within only 250 ms of network delay [34], the net-centric lifestyle based on robust, media-rich electronic workflow has strong potential of crossing borders and enabling innovations and virtual communities in ways that are not feasible or even foreseen today. Of particular interest we can also note the confluence of application-aware network services and mobile / rural broadband coverage, which may contribute towards acceleration of human development both in urban territories and communities that insofar have fallen behind the digital age.

## CONCLUSIONS AND FUTURE WORK

In this paper, we investigated the socioeconomic aspect of telecommuting. We have introduced the notion of Net-Centric Factor (NCF) and studied the links between intensity of telecommuting and feasible commute distances. Our formulated value of telecommute allowed us to show that network centricity allows remote workers to increase their effective settlement radius above and beyond the limits characteristic to legacy, commute-centric lifestyle.

Further, we have looked at economic incentives for telecommuters to increase their NCF and have found that such requests cannot be served with the currently available "best-effort" broadband infrastructure, thus pointing towards new network-based service opportunities.

Our work indicates that internet developers and international standards organizations have strong potential to develop new, high-margin and QoS-guaranteed consumer and business services. In a proposed extension of this work, we consider focusing on practical use-cases and simplification of existing architectures and orchestration abstractions down to practical, executable essentials.